\documentclass[10pt,amsmath,amssymb,pra,superscriptaddress, twocolumn]{revtex4}
\usepackage[dvipdfmx]{graphicx, color}
\usepackage{dcolumn}
\usepackage{bm}
\usepackage{booktabs}
\usepackage{color}
\newcounter{one}
\newcommand{\ket}[1]{| #1 \rangle}

\newcommand{\ca}[1]{{\cal #1}}

\newtheorem{theorem}{Theorem}
\newtheorem{lemma}{Lemma}
\newtheorem{definition}{Definition}
\newtheorem{definition2}{Definition}
\newtheorem{idea}{Basic Idea}

\def\QED{\mbox{\rule[0pt]{1.5ex}{1.5ex}}}

\def\endproof{\hspace*{\fill}~\QED\par\endtrivlist\unskip}

\newenvironment{proofof}[1]{\vspace*{5mm} \par \noindent
         {\bf Proof of #1:\hspace{2mm}}}{\endproof
}

\def\Label#1{\label{#1}\ [\ \text{#1}\ ]\ }
\def\Label{\label}

\newcommand{\affA}{Center for Emergent Matter Science (CEMS), RIKEN, Wako, Saitama 351-0198 Japan}
\newcommand{\affD}{Graduate School of Information Systems, The University of Electro-Communications, Chofu, Japan}

\begin{document}
\title{Regularized Boltzmann entropy determines macroscopic adiabatic accessibility}
\author{Hiroyasu Tajima}
\affiliation{\affA}
\author{Eyuri Wakakuwa}
\affiliation{\affD}

\begin{abstract}
How the thermodynamic entropy $S_{TD}$ is related to the Boltzmann entropy $S_{B}$ has been one of the central issues since the beginning of statistical mechanics.  Today, it is believed that the thermodynamic entropy $S_{TD}$ is equal to a function $\tilde{S}_{B}$ that is defined by regularizing the Boltzman entropy in order to ensure extensivity.  However, it is not known whether $\tilde{S}_{B}$ completely determines the possibility of a macroscopic adiabatic transformation in the same way as $S_{TD}$ does. In this paper, by formulating the possibility of  a macroscopic adiabatic transformations in terms of ``coarse-graining'' of quantum operations, we prove that $\tilde{S}_{B}$ provides a necessary and sufficient condition for the possibility of a macroscopic adiabatic transformation.
\end{abstract}
 
\maketitle

\section{Introduction}

Thermodynamics \cite{Fermi,Shimizu,Lieb} is one of the most successful phenomenologies in physics, and it has a huge application from chemical reactions \cite{Fermi} to black holes \cite{Bardeen}. 
The thermodynamic entropy $S_{TD}$ plays a central role in thermodynamics. It is a function of macroscopic variables such as $U$, $V$ and $N$, denoting the internal energy, the volume and the number of particles, respectively. As stated by the second law of thermodynamics, the thermodynamical entropy completely determines the macroscopic adiabatic accessibility, \cite{Lieb} i.e., 
\begin{align}
(U,V,N)&\prec_{ad}(U',V',N')\nonumber\\
&\Leftrightarrow S_{TD}[U,V,N]\le S_{TD}[U',V',N'].\label{st:secondlawOR}
\end{align}
where $(U,V,N)\prec_{ad}(U',V',N')$ means ``an adiabatic transformation from  a state $(U,V,N)$ to another state $(U',V',N')$ is possible''.

How the thermodynamic entropy $S_{TD}$ is related to the Boltzmann entropy $S_{B}$ has been one of the central issues since the beginning of statistical mechanics. 
The Boltzmann entropy $S_{B}$ is not equal to thermodynamic entropy $S_{TD}$ in general. 
For example, $S_{B}$ is not extensive in general,  while $S_{TD}$ is extensive.
Today, it is believed that $S_{TD}$ is equal to the regularized Boltzmann entropy $\tilde{S}_{B}$, which is defined in terms of $S_{B}$ as follows and is extensive by definition \cite{tasakistatistical}:
\begin{align}
&\tilde{S}_{B}[U,V,N]:=\lim_{X\rightarrow \infty}\frac{S_{B}[UX,VX,NX]}{X}.\Label{defSTD}
\end{align}
However, it is not known whether $\tilde{S}_{B}$ completely determines the macroscopic adiabatic accessibility in the same way as $S_{TD}$.
That is, it is not known whether the following statement holds:
\begin{align}
(U,V,N)&\prec_{ad}(U',V',N')\nonumber\\
&\Leftrightarrow \tilde{S}_{B}[U,V,N]\le \tilde{S}_{B}[U',V',N'].\label{st:secondlaw}
\end{align}

In the field of statistical mechanics, the forward implication of \eqref{st:secondlaw} has been proven for certain formulations of ``adiabatic operations,'' by only assuming that the limit \eqref{defSTD} exists and is a convex and increasing function for $U$, $V$ and $N$ \cite{Landau,Jarzynski,tasaki,Kurchan,Car1,Xiao,Sekimoto,tasaki15}.
However, the the backward implication of \eqref{st:secondlaw} is still left unproven \cite{Shimizu}.
Recently,  in the field of quantum information theory \cite{Horodecki,oneshot1,oneshot3,Egloff,Brandao,Car2,Popescu2014,Tajima2014,Weilenmann2015,review}, it has been succeeded in deriving detailed thermodynamic relations, which characterize possibility and impossibility of quantum state transformations by a set of restricted operations.
However, these conditions for the possibility of state transformations are represented not only by macroscopic parameters, but also by microscopic parameters. 
In this sence, their results are interpreted as the ``microscopic accessibility.''
This is in contrast to (\ref{st:secondlaw}), which is represented only by macroscopic parameters.

In this paper, we propose a coarse-graining approach to try the backward implication in \eqref{st:secondlaw}, and show that $\tilde{S}_{B}$ provides a necessary and sufficient condition for the possibility of a macroscopic adiabatic transformation, i.e., the macroscopic adiabatic accessibility.
Our results hold by only assuming that the limit \eqref{defSTD} exists and is a convex and strictly increasing function for $U$, $V$ and $N$.
Our resutls do not need  stronger assumptions, e.g, the i.i.d. feature.

This paper is organized as follows.
In the section \ref{s2}, we review basic and well-known concepts of thermodynamics and statistical mechanics. 
All contents in this section are well-known.
In the section \ref{s3}, we define the macroscopic adiabatic accessibility, based on our coarse-graining method.
In the section \ref{s4}, we give the main results of the present paper.
Finally, in the section \ref{s5}, we prove the main results. Several lemmas used in the section \ref{s5} are proven in Appendix.

\section{Preliminaries}
\label{s2}

In this section, we review basic concepts of thermodynamics and statistical mechanics. All contents in this section are well-known. See e.g. \cite{Shimizu,Lieb,Landau,tasakistatistical} for more details. 

In thermodynamics, an equilibrium state is represented by values of a set of macroscopic physical quantities such as $(U,V,N)$, where $U$ is the internal energy, $V$ is the volume of the system, and $N$ is the number of particles. In this paper, we consider cases where all these physical quantities are extensive,  the set includes the internal energy. 
We represent such a set of physical quantities in terms of vectors as $\vec{a}:=(a_{0}, a_{1},...,a_{L})$, where $L$ is a natural number, e.g., $L=2$ in the case of $(U,V,N)$. 
Here, for the simplicity of writing, we express the internal energy $U$ as $a_{0}$.
Since a macroscopic equilibrium state is uniquely determined by values of macroscopic physical quantities, we also represent an equilibrium state by $\vec{a}$. The thermodynamical entropy of an equilibrium state is uniquely determined as a function of $\vec{a}$, which we denote by $S_{TD}[\vec{a}]$.

The second law of thermodynamics is one of the most imporant assumptions in thermodynamics.
 It has several equivalent formulations such as the principle of maximum work, Clausius inequality and the law of entropy increase. The law of entropy increase determines the adiabatic accessibility, i.e., whether a macroscopic state can be transformed to another by an adiabatic operation. It is stated as follows:
\begin{align}
\vec{a}\prec_{ad}\vec{a}'\Leftrightarrow S_{TD}[\vec{a}]\le S_{TD}[\vec{a}'],
\end{align}
where $\vec{a}\prec_{ad}\vec{a}'$  means ``an adiabatic transformation $\vec{a}\rightarrow_{ad}\vec{a}$ is possible''.

Let us introduce the statistical mechanical counterpart of the thermodynamic equilibrium $\vec{a}$.
Since we are concerning a macroscopic limit, we describe a physical system by a Hilbert space ${\ca H}^{(X)}$ depending on a scaling parameter $X$.
The macroscopic limit is defined as the limit of $X\rightarrow\infty$. 
We assume that $X$ takes values in a set ${\mathcal X}={\mathbb N}$ or ${\mathcal X}={\mathbb R}^+$. 
For each $X\in{\ca X}$ and $l=0,\cdots,L$, we denote the set of the Hermitian operators on $\ca{H}^{(X)}$ as $\vec{A}^{(X)}:=(H^{(X)},A^{(X),[1]},...,A^{(X),[L]})$.
Then, the microcanonical state corresponding to an equilibrium state $\vec{a}$ is defined by
\begin{align}
\hat{\pi}^{(X)}_{\vec{a},\delta_{X}}:=\hat{\Pi}^{(X)}_{\vec{a},\delta_{X}}/D^{(X)}_{\vec{a},\delta_{X}},
\end{align}
where $\hat{\Pi}^{(X)}_{\vec{a},\delta_{X}}$ and $D^{(X)}_{\vec{a},\delta_{X}}$ are the projection and the dimension of 
the following $\ca{H}^{(X)}_{{\vec{a}},\delta_{X}}$, which is a subspace of  of $\ca{H}^{(X)}$:
\begin{align}
&\ca{H}^{(X)}_{\vec{a},\delta_{X}}:={\rm span}\left\{\ket{\psi}\in\ca{H}^{(X)}\right.\left|\:\exists\lambda^{[l]}\in[X(a^{[l]}-\delta_{X}),\nonumber \right.\nonumber\\
&\left. X(a^{[l]}+\delta_{X}))\text{ s.t. }A^{(X)[l]}\ket{\psi}=\lambda^{[l]}\ket{\psi}\text{ for $0\le l\le L$}\right\}.\Label{microcanonical}
\end{align}
The parameter $\delta_{X}$ is a positive function of $X$, which represents the negligible fluctuation of macroscopic quantities.
Since we are normalizing macroscopic observables as (\ref{microcanonical}), it is natural to assume that $\lim_{X\rightarrow\infty}\delta_X=0$.

To describe behavior of a system in the macroscopic limit ($X\rightarrow\infty$), we introduce an array $\{\delta_{X}\}_{X\in\ca{X}}$, and introduce an array of microcanonical states $\{\hat{\pi}^{(X)}_{\vec{a},\delta_{X}}\}_{X\in\ca{X}}$ for each $\vec{a}$ and $\{\delta_{X}\}_{X\in\ca{X}}$. 
Depending on the choice of $\{\delta_{X}\}_{X\in\ca{X}}$, there exists many arrays $\{\hat{\pi}^{(X)}_{\vec{a},\delta_{X}}\}_{X\in\ca{X}}$ corresponding to one $\vec{a}$.
Therefore, a thermodynamic equilibrium state $\vec{a}$ does not have a one-to-one correspondence with $\{\hat{\pi}^{(X)}_{\vec{a},\delta_{X}}\}_{X\in\ca{X}}$.
Let $\Delta$ denote the set of arrays $\{\delta_X\}$ such that $\lim_{X\rightarrow\infty}\delta_X=0$. Then $\vec{a}$ has a one-to-one correspondence with the following set of arrays of microcanonical state:
\begin{align}
\{\{\hat{\pi}^{(X)}_{\vec{a},\delta_{X}}\}_{X\in\ca{X}}|\{\delta_{X}\}_{X\in\ca{X}}\in\Delta\}.\label{setarray}
\end{align}
Since $\{\delta_{X}\}_{X\in\ca{X}}\in\Delta$ represents the range of ``negligible fluctuation'' of macroscopic physical quantities, any sequence in the set (\ref{setarray}) can be regarded as describing states that are ``macroscopically the same.''

The regularized Boltzmann entropy is defined as
\begin{align}
\tilde{S}_{B}[\vec{a}]:=\lim_{X\rightarrow \infty}\frac{1}{X}\log D^{(X)\downarrow}_{\vec{a}}.
\end{align}
where $D^{(X)\downarrow}_{\vec{a}}$ is the dimension of the Hilbert space $\ca{H}^{(X)\downarrow}_{{\vec{a}}}$ defined by
\begin{align}
&\ca{H}^{(X)}_{\vec{a}}:={\rm span}\left\{\ket{\psi}\in\ca{H}^{(X)}\right.\left|\:\exists\lambda^{[l]}\le Xa^{[l]},\right. \nonumber\\
&\left. \text{ s.t. }A^{(X)[l]}\ket{\psi}=\lambda^{[l]}\ket{\psi}\text{ for $0\le l\le L$}\right\},
\end{align}
With concrete calculations, it has been shown that there exists the limit $\tilde{S}_{B}$ as a convex and incresing function for each element of $\vec{a}$, in many physical systems, e.g., gases of particles with natural potentials including the van der Waars potential \cite{tasakistatistical}. 

\section{Formulation of macroscopic adiabatic accessibility in terms of coarse-graining}\label{s3}

In this section, we introduce a general method to formulate possibility of a macroscopic state transformation, by ``coarse-graining'' possibility of microscopic state transformations. We then formulate the macroscopic adiabatic accessibility, based on a coarse-graining of possibility of a microscopic state transformation by unital operations.

Our formulation is based on the following idea:

\begin{idea}
Suppose a microcanonical state $\pi^{(X)}_{\vec{a},\delta_{X}}$ is transformed by a quantum operation $\ca{E}_{X}$ to another microcanonical state $\pi^{(X)}_{\vec{a}',\delta'_{X}}$. From a macroscopic point of view, we observe that an equilibrium state $\vec{a}$ is transformed to another equilibrium state $\vec{a}'$, for any $\delta_{X}$ and $\delta'_{X}$ within the range of ``macroscopically negligible fluctuations''. Therefore, we could say that an equilibrium state $\vec{a}$ can be transformed to another equilibrium state $\vec{a}'$ if, for any macroscopically negligible $\delta_{X}$ and a $\delta'_{X}$, a state $\pi^{(X)}_{\vec{a},\delta_{X}}$ can be transformed to $\pi^{(X)}_{\vec{a}',\delta'_{X}}$.
\end{idea}

\begin{figure}[t]
\begin{center}
\includegraphics[scale=1.05,clip]{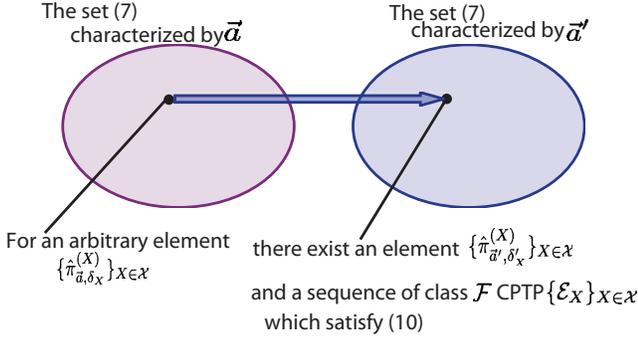}
\end{center}
\caption{Schematic diagram of Definition \ref{def:ad}}\Label{Fig2}
\end{figure}

Based on this idea and the concept of an array of microcanonical states introduced in Section \ref{s2}, we propose a definition of possibility of a macroscopic state transformation as follows, where $\ca{F}$ denotes a class of quantum operations (Fig. \ref{Fig2}).

\begin{definition}\label{def:ad}
{\it A macroscopic transformation $\vec{a}\rightarrow_{\tilde{\ca{F}}}\vec{a}'$ is possible} if, for any $\{\delta_X\}_{X\in\ca{X}}\in\Delta$, there exists $\{\delta_X'\}_{X\in\ca{X}}\in\Delta$ and an array $\{\ca{E}_{X}\}_{X\in \ca{X}}$ such that
\begin{align}
\lim_{n\rightarrow\infty}\left\|\ca{E}_{X}(\hat{\pi}^{(X)}_{\vec{a},\delta_{X}})-\hat{\pi}^{(X)}_{\vec{a}', \vec{\delta}'_{X}}\right\|=0,\Label{D1}
\end{align}
and $\ca{E}_{X}$ is a CPTP map of class $\ca{F}$ on ${\mathcal S}({\mathcal H}^X)$ for all $X\in\ca{X}$. Here, $\|\rho-\sigma\|$ is the trace distance defined by $\|\rho-\sigma\|:=\frac{1}{2}{\rm Tr}|\rho-\sigma|$. We express as $\vec{a}\prec_{\tilde{F}}\vec{a}'$ that an adiabatic transformation $\vec{a}\rightarrow_{\tilde{F}}\vec{a}'$ is possible.
\end{definition}
Clearly, if $\vec{a}\prec_{\tilde{\ca{F}}}\vec{a}'$ and $\vec{a}'\prec_{\tilde{\ca{F}}}\vec{a}''$ hold, then $\vec{a}\prec_{\tilde{\ca{F}}}\vec{a}''$ also holds.

The above definition provides a formulation of various types of possibility of macroscopic state transformations, depending on $\ca{F}$. In this paper, we employ unital CPTP map for class  $\ca{F}$ to define possibility of adiabatic transformations. Here, a unital CPTP map is defined as a completely positive trace preserving map that maps the identity operator to the identity operator, i.e., $\ca{E}(\hat{1})=\hat{1}$. A unital map does not decrease the von Neumann entropy of an arbitrary quantum state\cite{openquantum}, that is, we have\begin{align}
S(\ca{E}(\rho))\ge S(\rho),\enskip \forall \rho\in\ca{S}(\ca{H}),
\end{align}
for all $\rho\in\ca{S}(\ca{H})$.
Because this feature is similar to the feature of the adiabatic transformation in thermodynamics that does not decrease the thermodynamic entropy $S_{TD}$, 
many researches have treated the unital operation as a quantum counterpart of the adiabatic transformation in thermodynamics \cite{tasaki15,Popescu2014,Tajima2014,Weilenmann2015}.

We then formulate the macroscopic adiabatic accessibility as follows:
\begin{definition}\label{def:ad2}
An adiabatic transformation $\vec{a}\rightarrow_{ad}\vec{a}'$ is possible if, for any $\{\delta_X\}_{X\in\ca{X}}\in\Delta$, there exists $\{\delta_X'\}_{X\in\ca{X}}\in\Delta$ and an array $\{\ca{E}_{X}\}_{X\in \ca{X}}$ such that
\begin{align}
\lim_{X\rightarrow\infty}\left\|\ca{E}_{X}(\hat{\pi}^{(X)}_{\vec{a},\delta_{X}})-\hat{\pi}^{(X)}_{\vec{a}', \vec{\delta}'_{X}}\right\|=0,\Label{D1-2}
\end{align}
and $\ca{E}_{X}$ is a unital CPTP map on ${\mathcal S}({\mathcal H}^X)$ for all $X\in\ca{X}$. We express as $\vec{a}\prec_{ad}\vec{a}'$ when an adiabatic transformation $\vec{a}\rightarrow_{ad}\vec{a}'$ is possible.
\end{definition}

\section{Main Results}\label{s4}

The main results of this paper are summarized as follows.

\begin{theorem}\label{Thm}
When $\tilde{S}_{B}[\vec{a}]$ exists and is a convex and strictly increasing function for each element of $\vec{a}$, the following holds for arbitrary $\vec{a}$ and $\vec{a}'$:
\begin{align}
\tilde{S}_{B}[\vec{a}]\leq \tilde{S}_{B}[\vec{a}']\Leftrightarrow\vec{a}\prec_{ad}\vec{a}'.\Label{T1a}
\end{align}
\end{theorem}

When $\tilde{S}_{B}[\vec{a}]$ does not exist and is a convex and strictly increasing function for each element of $\vec{a}$, Theorem \ref{Thm} does not necessarily hold.
Even in that case, the following theorem holds in general: 
\begin{theorem}\label{Thm2}
We have
\begin{align}
\overline{\tilde{S}_{B}}[\vec{a}]< \underline{\tilde{S}_{B}}[\vec{a}']&\Rightarrow\vec{a}\prec_{ad}\vec{a}',\Label{(14)}\\
\underline{\tilde{S}_{B}}[\vec{a}]\le \overline{\tilde{S}_{B}}[\vec{a}']&\Leftarrow\vec{a}\prec_{ad}\vec{a}',\Label{(15)}
\end{align}
where we define
\begin{align}
\overline{\tilde{S}_{B}}[\vec{a}]&:=\sup_{\{\delta_{X}\}_{X\in\ca{X}}\in\Delta}\limsup_{X\rightarrow\infty}\frac{1}{X}S_{B}[\vec{a};\delta_{X}],\\
\underline{\tilde{S}_{B}}[\vec{a}]&:=\sup_{\{\delta_{X}\}_{X\in\ca{X}}\in\Delta}\liminf_{X\rightarrow\infty}\frac{1}{X}S_{B}[\vec{a};\delta_{X}],\Label{eq:defsupsaa}
\end{align}
where,
\begin{align}
S_{B}[\vec{a};\delta_{X}]:=\log D^{(X)}_{\vec{a},\delta_{X}}.
\end{align}
\end{theorem}

Theorem \ref{Thm} states that $\tilde{S}_{B}$ provides a necessary and sufficient condition for the macroscopic adiabatic accessibility in the same way as $S_{TD}$ does, as long as $\tilde{S}_{B}[\vec{a}]$ exists and is a convex and strictly increasing function for each element of $\vec{a}$. Theorem \ref{Thm2} provides a similar relation for the case where $\tilde{S}_{B}[\vec{a}]$ does not exist as a convex and strictly increasing function for each element of $\vec{a}$. The macroscopic adiabatic accessibility is in this case characterized by two functions $\overline{\tilde{S}_{B}}[\vec{a}]$ and $\underline{\tilde{S}_{B}}[\vec{a}]$, not by a single function as in the case of thermodynamic theories. 

Our results do not depend on any microscopic parameters, including $\delta_{X}$ that we have introduced to define the generalized microcanonical state $\hat{\pi}_{\vec{a},\delta_{X}}$. This is in contrast to previous approaches from quantum information theory \cite{Horodecki,oneshot1,oneshot3,Egloff,Brandao,Car2,Popescu2014,Tajima2014,Weilenmann2015,review}, which treat ``microscopic accessibility.'' 

We emphasize that Theorem \ref{Thm} only needs to assume that $\tilde{S}_{B}[\vec{a}]$ exists and is a convex and strictly increasing function for each element of $\vec{a}$, i.e., we do not need the stronger assumptions including the i.i.d. assumption.
It is shown that  $\tilde{S}_{B}[\vec{a}]$ exists and is a convex and increasing function for each element of $\vec{a}$, in many physical systems, e.g., gases of particles with natural potentials including the van der Waars potential \cite{tasakistatistical}.
In the region where the phase transition does not occur, the regularized Boltzmann entropy $\tilde{S}_{B}[\vec{a}]$ is a strictly increasing function for each element of $\vec{a}$.
Therefore, at least in such a region, the regularized Boltzmann entropy determines the macroscopic adiabatic accessibility.

\section{Proofs}\label{s5}

\subsection{Proof of Theorem \ref{Thm}}\label{s5-1}
In the proof of Theorem \ref{Thm}, we use the following three lemmas:
\begin{lemma}\Label{l0}
For an arbitrary $\{\delta_{X}\}\in\Delta$, the following relation holds:
\begin{align}
\lim_{X\rightarrow\infty}\frac{1}{X}\log D^{(X)\downarrow}_{\vec{a}(1+\delta_{X})}=\tilde{S}_{B}[\vec{a}]
\end{align}
\end{lemma}
\begin{lemma}\Label{l1l1}
When the regularized Boltzmann entropy $\tilde{S}_{B}[\vec{a}]$ exists for $\vec{a}$ and is a strictly increasing function for each element of $\vec{a}$, there exists $\{\delta^{(0)}_{X,\vec{a}}\}_{X\in\ca{X}}\in\Delta$ such that
\begin{align}
\lim_{X\rightarrow\infty}\frac{1}{X}\log D^{(X)}_{\vec{a},\delta_{X}}=\tilde{S}_{B}[\vec{a}]
\end{align}
holds for an arbitrary $\{\delta_{X}\}_{X\in\ca{X}}\in\Delta$ satisfying $\delta^{(0)}_{X,\vec{a}}<\delta_{X}$ for an arbitrary $X$.
\end{lemma}

\begin{lemma}\label{l1.5}
Let us consider the case where the regularized Boltzmann entropy $\tilde{S}_{B}[\vec{a}]$ exists and is a strictly increasing function for each element of $\vec{a}$.
Let us take arbitrary $\vec{a}$ and $\vec{a}'$ such that $\tilde{S}_{B}[\vec{a}]=\tilde{S}_{B}[\vec{a}']$.
Let us take an array $\{\delta_{X}\}\in\Delta$, and let $A_{X}$ be the following set of real numbers:
\begin{align}
A_{X}:=\{\eta|D^{(X)\downarrow}_{\vec{a}(1-\delta_{X})}\le D^{(X)\downarrow}_{\vec{a}'(1+\eta)}\le D^{(X)\downarrow}_{\vec{a}(1+\delta_{X})}\}
\end{align}
Let $\ca{X}_{0}$ be the set of $X$ such that $A_{X}$ is not the empty set, and let be $\ca{X}_{1}$ the set of $X$ such that $A_{X}$ is the empty set.
Then, we define two real valued functions $\eta^{(+)}_{X}$ and $\eta^{(-)}_{X}$ of $X$ as follows:
\begin{eqnarray}
\eta^{(+)}_{X}
&=\left\{ \begin{array}{ll}
\frac{1}{3}\inf_{\eta\in A_{X}}\eta+\frac{2}{3}\sup_{\eta\in A_{X}}\eta & (X\in\ca{X}_{0}) \\
\eta_{X}+\frac{1}{X} & (X\in\ca{X}_{1}) \\
\end{array} \right.,\\
\eta^{(-)}_{X}
&=\left\{ \begin{array}{ll}
\frac{2}{3}\inf_{\eta\in A_{X}}\eta+\frac{1}{3}\sup_{\eta\in A_{X}}\eta& (X\in\ca{X}_{0}) \\
\eta_{X}-\frac{1}{X} & (X\in\ca{X}_{1}) \\
\end{array} \right.,
\end{eqnarray}
where $\eta_{X}$ is the following real valued function of $X\in\ca{X}_{1}$:
\begin{align}
\eta_{X}:=\sup\{\eta|D^{(X)\downarrow}_{\vec{a}'(1+\eta)}<D^{(X)\downarrow}_{\vec{a}(1-\delta_{X})}\}.
\end{align}
Then, $\eta^{(+)}_{X}=o(1)$ and $\eta^{(-)}_{X}=o(1)$ hold.
\end{lemma}
These lemmas are proven in Appendix \ref{A0}.

Hereafter, we refer to the set of $\{\delta_{X}\}_{X\in\ca{X}}\in\Delta$ such that $\delta^{(0)}_{X,\vec{a}}<\delta_{X}<O(1)$ for arbitrary $X$ as $\Delta^{(0)}_{\vec{a}}$.
Clearly, for an arbitrary $\{\delta_{X}\}_{X\in\ca{X}}\in\Delta^{(0)}_{\vec{a}}$, the following relation holds:
\begin{align}
\frac{1}{X}\log D^{(X)}_{\vec{a},\delta_{X}}=\tilde{S}_{B}[\vec{a}]+f(\vec{a},\delta_{X},X)\label{fdef}.
\end{align}
Here, $f(\vec{a},\delta_{X},X)$ is a real-valued function which converges to 0 at the limit of $X\rightarrow\infty$.

\begin{proofof}{Theorem \ref{Thm}}
The forward implication of Proposition \eqref{T1a} is proved as follows. In general, there exists a unital CPTP map $\ca{E}$ satisfying $\sigma=\ca{E}(\rho)$, if and only if $\rho$ and $\sigma$ satisfies the majorization relation $\rho\prec_{M}\sigma$ defined as
\begin{align}
\rho\prec_{M}\sigma\underset{\mathrm{def}}\Longleftrightarrow\sum^{m}_{k=1}p^{\downarrow}_{k}\ge \sum^{m}_{k=1}q^{\downarrow}_{k}\enskip\mbox{for any $m$.}\Label{major}
\end{align}
Here, $\{p^{\downarrow}_{k}\}_k$ and $\{q^{\downarrow}_{k}\}_k$ are eigenvalues of $\rho$ and $\sigma$, respectively, sorted in decreasing order \cite{review,Nielsentext}. Hence it suffices to show that, for any $\{\delta_X\}_{X\in\ca{X}}\in\Delta$, there exists another sequence $\{\delta'_X\}_{X\in\ca{X}}\in\Delta$ such that we have
\begin{align}
\hat{\pi}^{(X)}_{\vec{a},\delta_{X}}\prec_{M}\hat{\pi}^{(X)}_{\vec{a}',\delta'_{X}}.\label{keyofL111}
\end{align}
for any sufficiently large $X$.

We firstly show that \eqref{keyofL111} holds when $\tilde{S}_{B}[\vec{a}]<\tilde{S}_{B}[\vec{a}']$ holds.
Let us take  arrays $\{\delta_X\}\in \Delta$ and  $\{\delta_X''\} \in \Delta_{a}^{(0)}$  satisfying $\delta_{X}<\delta''_{X}$ for arbitrary $X$.
Then, because $D^{(X)}_{\vec{a},\delta_{X}}$ is an increasing function for $\delta_{X}$, 
\begin{align}
\frac{1}{X}\log D^{(X)}_{\vec{a},\delta_{X}}\le\frac{1}{X}\log D^{(X)}_{\vec{a},\delta''_{X}}
\end{align}
holds.
Because of Lemma \ref{l1l1}, the righthand-side  of the above converges to $\tilde{S}_{B}[\vec{a}]$ at the limit of $X\rightarrow\infty$.
Also, because of Lemma \ref{l1l1}, for an arbitrary array $\{\delta'_{X}\}\in\Delta^{(0)}_{\vec{a}'}$, $\frac{1}{X}\log D^{(X)}_{\vec{a}',\delta'_{X}}$ converges to $\tilde{S}_{B}[\vec{a}']$ at the limit of $X\rightarrow\infty$.
Therefore, because of $\tilde{S}_{B}[\vec{a}]<\tilde{S}_{B}[\vec{a}']$,
\begin{align}
\frac{1}{X}\log D^{(X)}_{\vec{a},\delta_{X}}\le\frac{1}{X}\log D^{(X)}_{\vec{a}',\delta_{X}'}
\end{align}
holds for sufficient large $X$.
Hence, \eqref{keyofL111} holds when $\tilde{S}_{B}[\vec{a}]<\tilde{S}_{B}[\vec{a}']$ holds.

Next, let us show that \eqref{keyofL111} holds when $\tilde{S}_{B}[\vec{a}]=\tilde{S}_{B}[\vec{a}']$.
For an arbitrary $\{\delta_{X}\}\in\Delta$, we define
\begin{align}
\delta'_{X}:=4\max\{|\eta^{(+)}_{X}|,|\eta^{(-)}_{X}|\}.
\end{align}
Clearly,  $\delta'_{X}>\sup_{\eta\in A_{X}}\eta$ and $\inf_{\eta\in A_{X}}\eta>-\delta'_{X}$ hold for any $X\in\ca{X}_{0}$.
Therefore, for any $X\in\ca{X}_{0}$,
\begin{align}
D^{(X)\downarrow}_{\vec{a}(1+\delta_{X})}&<D^{(X)\downarrow}_{\vec{a}'(1+\delta'_{X})}\Label{9.19.1}\\
D^{(X)\downarrow}_{\vec{a}'(1-\delta'_{X})}&<D^{(X)\downarrow}_{\vec{a}(1-\delta_{X})}\Label{9.19.2}
\end{align}
hold. 
The inequalities \eqref{9.19.1} and \eqref{9.19.2} also hold for any $X\in\ca{X}_{1}$, because $\delta'_{X}>\eta_{X}$ and  $\eta_{X}>-\delta'_{X}$ hold for any $X\in\ca{X}_{1}$.
 (Note that $\eta>\eta_{X}\Rightarrow D^{(X)\downarrow}_{\vec{a}'(1+\eta)}>D^{(X)\downarrow}_{\vec{a}(1-\delta_{X})}$ and $\eta_{X}>\eta\Rightarrow D^{(X)\downarrow}_{\vec{a}(1-\delta_{X})}> D^{(X)\downarrow}_{\vec{a}'(1+\eta)}$ hold for an arbitrary $X\in\ca{X}_{1}$, because $\eta$ satisfying $D^{(X)\downarrow}_{\vec{a}(1-\delta_{X})}\le D^{(X)\downarrow}_{\vec{a}'(1+\eta)}\le D^{(X)\downarrow}_{\vec{a}(1+\delta_{X})}$ does not exist for $X\in\ca{X}_{1}$.)

Because of Lemma \ref{l1.5}, $\delta'_{X}=o(1)$ also holds.
Hence, $\{\delta'_{X}\}\in\Delta$ satisfies $D^{(X)}_{\vec{a},\delta_{X}}<D^{(X)}_{\vec{a}',\delta'_{X}}$, because of $D^{(X)}_{\vec{a},\delta_{X}}=D^{(X)\downarrow}_{\vec{a}(1+\delta_{X})}-D^{(X)\downarrow}_{\vec{a}(1-\delta_{X})}$.
Therefore, \eqref{keyofL111} holds.

Finally, we prove the backward implication of Proposition \eqref{T1a} by showing that an adiabatic transformation $\vec{a}\rightarrow_{aq}\vec{a}'$ is {\it not} possible if $\delta s:=\tilde{S}_{B}[\vec{a}]-\tilde{S}_{B}[\vec{a}']>0$. Fix arbitrary $\{\delta_{X}\}_{X\in\ca{X}}\in\Delta^{(0)}_{\vec{a}}$ and $\{\delta'_{X}\}_{X\in\ca{X}}\in\Delta$. 
We also take $\{\delta''_{X}\}_{X\in\ca{X}}\in\Delta^{(0)}_{\vec{a}'}$ such that $\delta'_{X}<\delta''_{X}$ for arbitrary $X$.
From \eqref{fdef}, we have
\begin{align}
&\frac{1}{X}\log{D^{(X)}_{\vec{a},\delta_X}}-\frac{1}{X}\log{D^{(X)}_{\vec{a}',\delta_X''}}\nonumber\\
&=(\tilde{S}_{B}[\vec{a}]-\tilde{S}_{B}[\vec{a}'])+(f(\vec{a},\delta_{X},X)-f(\vec{a}',\delta''_{X},X))\nonumber\\
&=\delta s-\gamma_X
\end{align}
for any $X$, where we defined $\gamma_X:= f(\vec{a},\delta_{X},X)-f(\vec{a}',\delta''_{X},X)$.
Hence we have
\begin{align}
&\frac{1}{X}\log{D^{(X)}_{\vec{a},\delta_X}}-\frac{1}{X}\log{D^{(X)}_{\vec{a}',\delta_X'}}\nonumber\\
&\ge\frac{1}{X}\log{D^{(X)}_{\vec{a},\delta_X}}-\frac{1}{X}\log{D^{(X)}_{\vec{a}',\delta_X''}}>\frac{\delta s}{2}
\end{align}
for any sufficiently large $X$, which leads to
\begin{eqnarray}
\frac{1}{D^{(X)}_{\vec{a}',\delta'_{X}}}>\frac{e^{-X\delta{s}/2}}{D^{(X)}_{\vec{a}',\delta'_{X}}}>\frac{1}{D^{(X)}_{\vec{a},\delta_{X}}}.\Label{eq:ddddcc}
\end{eqnarray}
Note that $\lim_{X\rightarrow\infty}\gamma_X=0$ follows from $\lim_{X\rightarrow\infty}f(\vec{a},\delta_{X},X)=0$ and $\lim_{X\rightarrow\infty}f(\vec{a}',\delta''_{X},X)=0$. 
Let $P^{(X)}_{\vec{a}',\delta'_{X}}$ be the projection onto ${\rm supp}[{\hat{\pi}^{(X)}_{\vec{a}',\delta'_{X}}}]$, and define a unital CPTP map $\ca{P}^{(X)}_{\vec{a}',\delta'_{X}}$ by
\begin{eqnarray}
\ca{P}^{(X)}_{\vec{a}',\delta'_{X}}(\rho)=P^{(X)}_{\vec{a}',\delta'_{X}}\rho P^{(X)}_{\vec{a}',\delta'_{X}}+(I-P^{(X)}_{\vec{a}',\delta'_{X}})\rho(I-P^{(X)}_{\vec{a}',\delta'_{X}}).\nonumber
\end{eqnarray}
Let $\ca{E}_{X}':=\ca{P}^{(X)}_{\vec{a}',\delta'_{X}}\circ\ca{E}_{X}$ for any unital CPTP map $\ca{E}_{X}$. Since $\ca{E}_{X}'$ is a untal CPTP map as well, we have
\begin{eqnarray}
\hat{\pi}^{(X)}_{\vec{a},\delta_{X}}\prec_{M}\ca{E}_{X}(\hat{\pi}^{(X)}_{\vec{a},\delta_{X}})\prec_{M}\ca{E}_{X}'(\hat{\pi}^{(X)}_{\vec{a},\delta_{X}}).
\end{eqnarray}
Consequently, all eigenvalues $p_i\:(i=1,2,\cdots)$ of $\ca{E}_{X}'(\hat{\pi}^{(X)}_{\vec{a},\delta_{X}})$ satisfies
\begin{eqnarray}
\frac{1}{D^{(X)}_{\vec{a},\delta_{X}}}\ge p_{i}.\Label{eq:taskact2}
\end{eqnarray}
From \eqref{eq:ddddcc} and \eqref{eq:taskact2}, we obtain
\begin{align}
\frac{1}{D^{(X)}_{\vec{a}',\delta'_{X}}}>\frac{e^{-X\delta s/2}}{D^{(X)}_{\vec{a}',\delta'_{X}}}\ge p_{i}.\label{eq:kanedaichi}
\end{align}

The distance between $\ca{E}_{X}(\hat{\pi}^{(X)}_{\vec{a},\delta_{X}})$ and $\hat{\pi}^{(X)}_{\vec{a}',\delta'_{X}}$ is then bounded as follows. Due to the monotonicity of the trace distance, we have
\begin{eqnarray}
&&\left\|\ca{E}_{X}(\hat{\pi}^{(X)}_{\vec{a},\delta_{X}})-\hat{\pi}^{(X)}_{\vec{a}',\delta'_{X}}\right\|\ge\|\ca{E}_{X}'(\hat{\pi}^{(X)}_{\vec{a},\delta_{X}})-\hat{\pi}^{(X)}_{\vec{a}',\delta'_{X}}\|\nonumber\\
&=&\frac{1}{2}\sum^{D^{(X)}_{\vec{a}',\delta'_{X}}}_{i=1}\left|p_{i}-\frac{1}{D^{(X)}_{\vec{a}',\delta'_{X}}}\right|+\frac{1}{2}\sum_{i>D^{(X)}_{\vec{a}',\delta'_{X}}}p_{i}\nonumber\\
&\geq&\frac{1}{2}\sum^{D^{(X)}_{\vec{a}',\delta'_{X}}}_{i=1}\left|p_{i}-\frac{1}{D^{(X)}_{\vec{a}',\delta'_{X}}}\right|\nonumber
\\&\geq&\frac{1}{2}D^{(X)}_{\vec{a}',\delta'_{X}}\left(\frac{1}{D^{(X)}_{\vec{a}',\delta'_{X}}}-\frac{e^{-X\delta{s}/2}}{D^{(X)}_{\vec{a}',\delta'_{X}}}
\right)\nonumber\\
&=&\frac{1}{2}\left(1-e^{-X\delta{s}/2}\right).
\end{eqnarray}
Here, we defined $p_i\;(i=1,\cdots,{D^{(X)}_{\vec{a}',\delta'_{X}}})$ as eigenvalues of $\ca{E}_{X}'(\hat{\pi}^{(X)}_{\vec{a},\delta_{X}})$ on the support of ${\hat{\pi}^{(X)}_{\vec{a}',\delta'_{X}}}$. Thus we obtain
\begin{align}
\left\|\ca{E}_{X}(\hat{\pi}^{(X)}_{\vec{a},\delta_{X}})-\hat{\pi}^{(X)}_{\vec{a}',\delta'_{X}}\right\|\ge\frac{1}{3}\label{eq:mikitaka}
\end{align}
for any sufficiently large $X$ and any unital CPTP map $\ca{E}_{X}$, which contradicts \eqref{D1-2}. Therefore, an adiabatic transformation $\vec{a}\rightarrow_{aq}\vec{a}'$ is not possible.
\end{proofof}

\subsection{Proof of Theorem \ref{Thm2}}

We first introduce a generalized adiabatic process.
\begin{definition}\label{den:dannnetsukanou2}
Suppose $\{\delta_{X}\}_{X\in\ca{X}},\{\delta'_{X}\}_{X\in\ca{X}}\in\Delta$.
An adiabatic transformation $(\vec{a},\{\delta_{X}\}_{X\in\ca{X}})\rightarrow_{ad}(\vec{a}',\{\delta'_{X}\}_{X\in\ca{X}})$ is possible if there exists a set $\{\ca{E}_X\}$ of unital CPTP maps $\ca{E}_X$ on $\ca{S}(\ca{H}^{(X)})$ that satisfies
\begin{align}
\lim_{n\rightarrow\infty}\left\|\ca{E}_{X}\left(\hat{\pi}^{(X)}_{\vec{a},\delta_X}\right)-\hat{\pi}^{(X)}_{\vec{a}',\delta_X'}\right\|=0.\label{eq:defgenad}
\end{align}
\end{definition}
Definition \ref{def:ad2} is then reformulated as follows:
\begin{definition2}\label{den:dannnetsukanou1'}
An adiabatic transformation $\vec{a}\rightarrow_{ad}\vec{a}'$ is possible if, for any $\{\delta_{X}\}_{X\in\ca{X}}\in\Delta$, there exists another sequence $\{\delta'_{X}\}_{X\in\ca{X}}\in\Delta$, such that an adiabatic transformation $(\vec{a},\{\delta_{X}\}_{X\in\ca{X}})\rightarrow_{ad}(\vec{a}',\{\delta'_{X}\}_{X\in\ca{X}})$ is possible.
\end{definition2}
Theorem \ref{Thm2} is proved by using the following lemma regarding the possibility and impossibility of generalized adiabatic processes. A proof is given in Appendix \ref{AppendixB}.

\begin{lemma}\label{Thm22}
Let us define
\begin{eqnarray}
&&{\overline s}_{\vec{a},\{\delta_{X}\}_{X\in\ca{X}}}:=\limsup_{X\rightarrow\infty}\frac{1}{X}S_{B}[\vec{a},\delta_{X}]\label{sup}\\
&&{\underline s}_{\vec{a},\{\delta_{X}\}_{X\in\ca{X}}}:=\liminf_{X\rightarrow\infty}\frac{1}{X}S_{B}[\vec{a},\delta_{X}]\label{inf}
\end{eqnarray}
for arbitrary $\vec{a}$ and $\{\delta_{X}\}_{X\in\ca{X}}\in\Delta$.
Suppose $\{\delta_{X}\}_{X\in\ca{X}},\{\delta_{X}'\}_{X\in\ca{X}}\in\Delta$.
Then, an adiabatic transformation $(\vec{a},\{\delta_{X}\}_{X\in\ca{X}})\rightarrow_{ad}(\vec{a}',\{\delta'_{X}\}_{X\in\ca{X}})$ is possible if ${\overline s}_{\vec{a},\{\delta_{X}\}_{X\in\ca{X}}}< {\underline s}_{\vec{a}',\{\delta'_{X}\}_{X\in\ca{X}}}$. Conversely, an adiabatic transformation $(\vec{a},\{\delta_{X}\}_{X\in\ca{X}})\rightarrow_{ad}(\vec{a}',\{\delta'_{X}\}_{X\in\ca{X}})$ is possible only if ${\underline s}_{\vec{a},\{\delta_{X}\}_{X\in\ca{X}}}\leq {\overline s}_{\vec{a}',\{\delta'_{X}\}_{X\in\ca{X}}}$.
\end{lemma}
Theorem \ref{Thm2} is then proved as follows.

\begin{proofof}{Theorem \ref{Thm2}}
The relation \eqref{(14)} is proved as follows. Due to
\begin{align}
\overline{\tilde{S}_{B}}[\vec{a}]=\sup_{\{\delta_{X}\}_{X\in\ca{X}}\in\Delta}{\overline s}_{\vec{a},\{\delta_{X}\}_{X\in\ca{X}}},\nonumber\\
\underline{\tilde{S}_{B}}[\vec{a}]=\sup_{\{\delta_{X}\}_{X\in\ca{X}}\in\Delta}{\underline s}_{\vec{a},\{\delta_{X}\}_{X\in\ca{X}}},
\end{align}
we have
\begin{eqnarray}
&&\forall\{\delta_{X}\}_{X\in\ca{X}}\in\Delta\:;\;\;{\overline s}_{\vec{a},\{\delta_{X}\}_{X\in\ca{X}}}\leq \overline{\tilde{S}_{B}}[\vec{a}],\nonumber\\
&&\forall\epsilon>0,\exists\{\delta'_{X}\}_{X\in\ca{X}}\in\Delta\:;\;\;\underline{\tilde{S}_{B}}[\vec{a}']-\epsilon\leq{\underline s}_{\vec{a}',\{\delta'_{X}\}_{X\in\ca{X}}}.\;\;\;\;\;\label{eq:killerqueen}
\end{eqnarray}
By choosing $\epsilon=(\underline{\tilde{S}_{B}}[\vec{a}']-\overline{\tilde{S}_{B}}[\vec{a}])/2>0$, it follows that for any $\{\delta_{X}\}_{X\in\ca{X}}$, there exists $\{\delta'_{X}\}_{X\in\ca{X}}$ such that
\begin{eqnarray}
{\overline s}_{\vec{a},\{\delta_{X}\}_{X\in\ca{X}}}\leq \overline{\tilde{S}_{B}}[\vec{a}]<\frac{\underline{\tilde{S}_{B}}[\vec{a}']+\overline{\tilde{S}_{B}}[\vec{a}]}{2}\leq{\underline s}_{\vec{a}',\{\delta'_{X}\}_{X\in\ca{X}}},
\end{eqnarray}
due to Lemma \ref{Thm22}.
Thus an adiabatic transformation $a\rightarrow a'$ is possible. To prove \eqref{(15)}, we assume that $\underline{\tilde{S}_{B}}[\vec{a}]>\overline{\tilde{S}_{B}}[\vec{a}']$. Equivalently to (\ref{eq:killerqueen}), we have
\begin{eqnarray}
&&\forall\epsilon>0,\exists\{\delta_{X}\}_{X\in\ca{X}}\in\Delta\:;\;\;\underline{\tilde{S}_{B}}[\vec{a}]-\epsilon\leq{\underline s}_{\vec{a},\{\delta_{X}\}_{X\in\ca{X}}}\nonumber\\
&&\forall\{\delta'_{X}\}_{X\in\ca{X}}\in\Delta\:;\;\;{\overline s}_{\vec{a}',\{\delta'_{X}\}_{X\in\ca{X}}}\leq \overline{\tilde{S}_{B}}[\vec{a}'].
\end{eqnarray}
By choosing $\epsilon=(\underline{\tilde{S}_{B}}[\vec{a}]-\overline{\tilde{S}_{B}}[\vec{a}'])/2>0$, we find that there exists $\{\delta_{X}\}_{X\in\ca{X}}$, such that for any $\{\delta'_{X}\}_{X\in\ca{X}}$ we have
\begin{eqnarray}
{\underline s}_{\vec{a},\{\delta_{X}\}_{X\in\ca{X}}}\geq \frac{\underline{\tilde{S}_{B}}[\vec{a}]+\overline{\tilde{S}_{B}}[\vec{a}']}{2}>\overline{\tilde{S}_{B}}[\vec{a}']\geq{\overline s}_{\vec{a}',\{\delta'_{X}\}_{X\in\ca{X}}},
\end{eqnarray}
due to Lemma \ref{Thm22}.
Thus an adiabatic transformation $\vec{a}\rightarrow_{ad}\vec{a}'$ is not possible, which completes the proof.
\end{proofof}

\hfill

\section{Conclusion}

We have proved that the regularized Boltzmann entropy $\tilde{S}_{B}$ gives a necessary and sufficient condition for the macroscopic accessibility, in the same way as the thermodynamic entropy $S_{TD}$ does, by formulating adiabatic operations in terms of a ``coarse-grained'' unital operations.
The result is applicable for any physical system in which $\tilde{S}_{B}[\vec{a}]$ exists as a continous and strictly increasing function for each element of $\vec{a}$.
It is shown that  $\tilde{S}_{B}[\vec{a}]$ exists and is a convex and increasing function for each element of $\vec{a}$, in many physical systems, e.g., gases of particles with natural potentials including the van der Waars potential \cite{tasakistatistical}.
In the region where the phase transition does not occur, the regularized Boltzmann entropy $\tilde{S}_{B}[\vec{a}]$ is a strictly increasing function for each element of $\vec{a}$.
Therefore, at least in such a region, the regularized Boltzmann entropy determines the macroscopic adiabatic accessibility.

We also proved that a similar relation holds in general, even for systems in which $\tilde{S}_{B}[\vec{a}]$ may not exist as a convex and strictly increasing function for each element of $\vec{a}$.
Possibility of an adiabatic transformation is in this case characterized by two functions $\overline{\tilde{S}_{B}}[\vec{a}]$ and $\underline{\tilde{S}_{B}}[\vec{a}]$, not by a single function as in the case of thermodynamic theories.

We emphasize that Theorem \ref{Thm} do not need the stronger assumptions including the i.i.d. assumption.
Also, our results do not depend on any microscopic parameters, including $\delta_{X}$ that we have introduced to define the generalized microcanonical state $\hat{\pi}_{\vec{a},\delta_{X}}$. This is in contrast to previous approaches from quantum information theory \cite{Horodecki,oneshot1,oneshot3,Egloff,Brandao,Car2,Popescu2014,Tajima2014,Weilenmann2015,review} treating ``microscopic accessibility'', in which convertibility of states are characterized by functions that depends on microscopic parameters.

Our method, presented in Definition \ref{def:ad2} to formulate possibility of macroscopic transformations in terms coarse-graining, is applicable as well for classes of quantum operations other than unital operations. To find a class of operations that genuinely described adiabatic operations, and to see whether the same results as Theorem \ref{Thm} and \ref{Thm2} hold for that class, are left as future works.

\textit{Acknowledgments:}
HT is grateful to Prof. Hal Tasaki, Prof. Gen Kimura, Dr. Yu Watanabe, Dr. Sho Sugiura, Dr. Kiyoshi Kanazawa,  Dr. Jun'ichi Ozaki, Dr. Ryoto Sawada, Dr. Sosuke Ito and Yohei Morikuni for helpful comments.

\renewcommand{\refname}{\vspace{-1cm}}

\widetext

\appendix

\section{Proof of Lemmas \ref{l0}, \ref{l1l1} \ref{l1.5}}\label{A0}

\begin{proofof}{Lemma \ref{l0}}
Because $\delta_{X}=o(1)$ and $D^{(X)\downarrow}_{\vec{a}}$ is increasing for each element of $\vec{a}$, the following inequality holds for an arbitrary $\epsilon>0$ and sufficiently large $X$:
\begin{align}
\frac{1}{X}\log D^{(X)\downarrow}_{\vec{a}(1-\epsilon)}<\frac{1}{X}\log D^{(X)\downarrow}_{\vec{a}(1+\delta_{X})}<\frac{1}{X}\log D^{(X)\downarrow}_{\vec{a}(1+\epsilon)}.
\end{align}
Therefore, for an arbitrary $\epsilon>0$,
\begin{align}
\tilde{S}_{B}[\vec{a}(1-\epsilon)]\le\liminf\frac{1}{X}\log D^{(X)\downarrow}_{\vec{a}(1+\delta_{X})}\le\limsup\frac{1}{X}\log D^{(X)\downarrow}_{\vec{a}(1+\delta_{X})}<\tilde{S}_{B}[\vec{a}(1+\epsilon)]
\end{align}
holds.
Because $\tilde{S}_{B}$ is continuous, Lemma \ref{l0} holds.
\end{proofof}

\begin{proofof}{Lemma \ref{l1l1}}
For an arbitrary $\{\delta_{X}\}\in\Delta$, 
\begin{align}
\frac{1}{X}\log D^{(X)}_{\vec{a},\delta_{X}}&=\frac{1}{X}\log(D^{(X)\downarrow}_{\vec{a}(1+\delta_{X})}-D^{(X)\downarrow}_{\vec{a}(1-\delta_{X})})\nonumber\\
&=\frac{1}{X}\log D^{(X)\downarrow}_{\vec{a}(1+\delta_{X})}+\frac{1}{X}\log\left(1-\frac{D^{(X)\downarrow}_{\vec{a}(1-\delta_{X})}}{D^{(X)\downarrow}_{\vec{a}(1+\delta_{X})}}\right).\label{secondpart}
\end{align}
Because of Lemma \ref{l0}, the first part of the right-hand side converges to $\tilde{S}_{B}[\vec{a}]$ at the limit of $X$.
Therefore, we only need to show that the second part converges to 0 for an arbitrary$\{\delta_{X}\}\in\Delta$ such that $\delta^{(0)}_{X,\vec{a}}<\delta_{X}$ for a $\{\delta^{(0)}_{X,\vec{a}}\}\in\Delta$.
Because $D^{(X)\downarrow}_{\vec{a}}$ is increasing for each element of $\vec{a}$, it is sufficient to show that the second part vanishes for an $\{\delta^{(0)}_{X,\vec{a}}\}\in\Delta$.

For an arbitrary $\vec{a}$, the relation $\frac{1}{X}\log D^{(X)\downarrow}_{\vec{a}}\rightarrow\tilde{S}_{B}[\vec{a}]$ holds.
Therefore, there exists a function $\gamma_{\vec{a},X}$ satisfying $\lim_{X\rightarrow\infty}\gamma_{\vec{a},X}=0$, and 
\begin{align}
\frac{1}{X}\log D^{(X)\downarrow}_{\vec{a}}=\tilde{S}_{B}[\vec{a}]+\gamma_{\vec{a},X}
\end{align}
holds.
Therefore, we can transform the second part of \eqref{secondpart} as follows:
\begin{align}
\frac{1}{X}\log(1-e^{X(\tilde{S}_{B}[\vec{a}(1-\delta_{X})]+\gamma_{\vec{a}(1-\delta_{X}),X}-\tilde{S}_{B}[\vec{a}(1+\delta_{X})]-\gamma_{\vec{a}(1+\delta_{X}),X})})
\end{align}
Because
\begin{align}
0\ge\frac{1}{X}\log(1-e^{Xg(X)})\ge\frac{1}{X}\log(1-e^{-1/\sqrt{X}})\rightarrow 0
\end{align}
holds for an arbitrary $g(X)\le-\frac{1}{\sqrt{X}}$,  we only need to show that 
\begin{align}
\tilde{S}_{B}[\vec{a}(1-\delta_{X})]+\gamma_{\vec{a}(1-\delta_{X}),X}-\tilde{S}_{B}[\vec{a}(1+\delta_{X})]-\gamma_{\vec{a}(1+\delta_{X}),X}\le-\frac{1}{\sqrt{X}}\label{l2key}
\end{align}
holds for an array $\{\delta^{(0)}_{X,\vec{a}}\}\in\Delta$.
Because $\tilde{S}_{B}[\vec{a}]$ is strictly increasing for each element of $\vec{a}$, for an arbitrary $\epsilon>0$, there exists $X_{\epsilon}>0$ such that
\begin{align}
\tilde{S}_{B}[\vec{a}(1-\epsilon)]+\gamma_{\vec{a}(1-\epsilon),X}-\tilde{S}_{B}[\vec{a}(1+\epsilon)]-\gamma_{\vec{a}(1+\epsilon),X}\le-\frac{1}{\sqrt{X}}
\end{align}
holds for  arbitrary $X>X_{\epsilon}$.
Therefore, the following $\delta^{(0)}_{X,\vec{a}}\in\Delta$ satisfies \eqref{l2key}:
\begin{align}
X_{1}&=X_{\epsilon_{1}},\\
X_{m}&=\max\{X_{\epsilon_{m}},X_{m-1}+1\},\\
\delta^{(0)}_{X,\vec{a}}&=\epsilon_{m} \enskip\mbox{for} \enskip X_{m+1}\ge X\ge X_{m}
\end{align}
where $\{\epsilon_{m}\}$ is an array of positive numbers such that $\epsilon_{m}\rightarrow_{m\rightarrow\infty}0$.
\end{proofof}

\begin{proofof}{Lemma \ref{l1.5}}
We firstly consider the case that $\ca{X}_{1}$ is bounded.
In this case, we only need to consider $X\in \ca{X}_{0}$.
By definition, for an arbitrary $X\in \ca{X}_{0}$, 
\begin{align}
\frac{1}{X}\log D^{(X)\downarrow}_{\vec{a}(1-\delta_{X})}&\le \frac{1}{X}\log D^{(X)\downarrow}_{\vec{a}'(1+\eta^{(-)}_{X})}\nonumber\\
&\le \frac{1}{X}\log D^{(X)\downarrow}_{\vec{a}'(1+\eta^{(+)}_{X})}\le \frac{1}{X}\log D^{(X)\downarrow}_{\vec{a}(1+\delta_{X})}
\end{align}
Because of Lemma \ref{l0} and $\tilde{S}_{B}[\vec{a}]=\tilde{S}_{B}[\vec{a}']$, the relation $\frac{1}{X}\log D^{(X)\downarrow}_{\vec{a}'(1+\eta^{(+)}_{X})}\rightarrow\tilde{S}_{B}[\vec{a}']$ holds.
If $\eta^{(+)}_{X}=o(1)$ would not hold, there exists a subsequence of $\{\eta^{(+)}_{X}\}$ such that $\eta^{(+)}_{X}=\epsilon+\tilde{\eta_{X}}$ for a real number $\epsilon\ne0$ and $\tilde{\eta}_{X}=o(1)$.
Because of Lemma 1, the subsequence satisfies $\frac{1}{X}\log D^{(X)\downarrow}_{\vec{a}'(1+\eta^{(+)}_{X})}\rightarrow\tilde{S}_{B}[\vec{a}'(1+\epsilon)]$.
It is a contradiction, and therefore $\eta^{(+)}_{X}=o(1)$ holds.
We can show $\eta^{(-)}_{X}=o(1)$ in the same manner.

Next, we consider the case where $\ca{X}_{1}$ is unbounded.
For the part of $X\in \ca{X}_{0}$, $\eta^{(+)}_{X}=o(1)$ and $\eta^{(-)}_{X}=o(1)$ has been shown.
Thus we only need to show $\eta^{(+)}_{X}=o(1)$ and $\eta^{(-)}_{X}=o(1)$ for $X\in \ca{X}_{1}$.
It is sufficient to show $\eta_{X}=o(1)$ for $X\in \ca{X}_{1}$.

For $X\in \ca{X}_{1}$, 
\begin{align}
\frac{1}{X}\log D^{(X)\downarrow}_{\vec{a}'(1+\eta_{X}-\frac{1}{X})}&\le \frac{1}{X}\log D^{(X)\downarrow}_{\vec{a}(1-\delta_{X})}\label{kuhu1}\\
\frac{1}{X}\log D^{(X)\downarrow}_{\vec{a}(1+\delta_{X})}&\le \frac{1}{X}\log D^{(X)\downarrow}_{\vec{a}'(1+\eta_{X}+\frac{1}{X})}\label{kuhu2}
\end{align}
holds, where \eqref{kuhu2} holds because there is no $\eta$ satisfying   $D^{(X)\downarrow}_{\vec{a}(1-\delta_{X})}\le D^{(X)\downarrow}_{\vec{a}'(1+\eta)}\le D^{(X)\downarrow}_{\vec{a}(1+\delta_{X})}$ for $X\in \ca{X}_{1}$. 
If $\eta_{X}=o(1)$ would not hold, either 
\begin{align}
\overline{\eta}&:=\limsup_{X\rightarrow\infty} \eta_{X}>0\label{kuhu3}
\end{align}
or
\begin{align}
\underline{\eta}&:=\liminf_{X\rightarrow\infty} \eta_{X}<0\label{kuhu4}
\end{align}
holds.
Note that $\eta_{X}=\underline{\eta}+o(1)$ holds for a subsequence of $\{\eta_{X}\}_{X\in\ca{X}_{1}}$, and that $\eta_{X}=\overline{\eta}+o(1)$ holds for another subsequence.
Because of Lemma 1, for each subsequences, the following relation holds respectively:
\begin{align}
\frac{1}{X}\log D^{(X)\downarrow}_{\vec{a}'(1+\eta_{X}-\frac{1}{X})}&\rightarrow \tilde{S}_{B}[\vec{a}(1+\overline{\eta})]\\
\frac{1}{X}\log D^{(X)\downarrow}_{\vec{a}'(1+\eta_{X}+\frac{1}{X})}&\rightarrow \tilde{S}_{B}[\vec{a}(1+\underline{\eta})].
\end{align}
Therefore, because of \eqref{kuhu1} and \eqref{kuhu2},
\begin{align}
\tilde{S}_{B}[\vec{a}'(1+\overline{\eta})]\le \tilde{S}_{B}[\vec{a}'],\\
\tilde{S}_{B}[\vec{a}'(1+\underline{\eta})]\ge \tilde{S}_{B}[\vec{a}'].
\end{align}
Hence we have $\overline{\eta}=\underline{\eta}=0$. It contradicts the claim that either \eqref{kuhu3} or \eqref{kuhu4} holds.
\end{proofof}

\section{Proof of Lemma \ref{Thm22}}\label{AppendixB}

There exists a unital map $\mathcal E$ satisfying ${\mathcal E}(\rho)=\sigma$, if and only if $\rho\prec\sigma$ holds, i.e., $\sigma$ is majorized by $\rho$ (see \eqref{major} for the definition). Thus we prove the first statement of Lemma \ref{Thm22} by showing that if $\overline{s}<\underline{s}$, we have 
\begin{align}
\hat{\pi}^{(X)}_{\vec{a},\delta_X}\prec\hat{\pi}^{(X)}_{\vec{a}',\delta_X'}\label{keyofL12}
\end{align}
for any $\{\delta_X\}_{X\in\ca{X}},\{\delta'_X\}_{X\in\ca{X}}\in\Delta$ and sufficiently large $X$. Define
\begin{eqnarray}
&&{\overline\gamma}_{\vec{a},\{\delta_X\}_{X\in\ca{X}},X}:=\max\left\{\frac{1}{X}S(\hat{\pi}^{(X)}_{\vec{a},\delta_X})-{\overline s}_{\vec{a},\{\delta_X\}_{X\in\ca{X}}},\;0\right\},\nonumber\\
&&{\underline\gamma}_{\vec{a}\{\delta_X\}_{X\in\ca{X}},X}:= -\min\left\{\frac{1}{X}S(\hat{\pi}^{(X)}_{\vec{a},\delta_X})-{\underline s}_{\vec{a}',\{\delta'_X\}_{X\in\ca{X}}},\;0\right\}\nonumber
\end{eqnarray}
for any $\{\delta_X\}_{X\in\ca{X}}\in\Delta$. By definition, we have
\begin{align}
&{\underline s}_{\vec{a},\{\delta_X\}_{X\in\ca{X}}}-{\underline\gamma}_{\vec{a},\{\delta_X\}_{X\in\ca{X}},X}\nonumber\\
&\leq \frac{S(\hat{\pi}^{(X)}_{\vec{a},\delta_X})}{X}\leq {\overline s}_{\vec{a},\{\delta_X\}_{X\in\ca{X}}}+{\overline\gamma}_{\vec{a},\{\delta_X\}_{X\in\ca{X}},X}\label{eq:boundpmmm}
\end{align}
for any $\vec{a}$ and $X$, as well as
\begin{align}
\lim_{X\rightarrow\infty}{\overline\gamma}_{\vec{a},\{\delta_X\}_{X\in\ca{X}},X}=\lim_{X\rightarrow\infty}{\underline\gamma}_{\vec{a},\{\delta_X\}_{X\in\ca{X}},X}=0\label{eq:gammaconv}
\end{align}
due to the definitions of ${\overline s}_{\vec{a},\{\delta_X\}_{X\in\ca{X}}}$ and ${\underline s}_{\vec{a},\{\delta_X\}_{X\in\ca{X}}}$. Hence we have
\begin{align}
&\frac{1}{X}\log{D^{(X)}_{\vec{a},\delta_X}}=\frac{1}{X}S(\hat{\pi}^{(X)}_{\vec{a},\delta_X})\leq{\overline s}_{\vec{a},\{\delta_{X}\}_{X\in\ca{X}}}+{\overline\gamma}_{\vec{a},\{\delta_X\}_{X\in\ca{X}},X},\nonumber\\
&\frac{1}{X}\log{D^{(X)}_{\vec{a}',\delta_X'}}=\frac{1}{X}S(\hat{\pi}^{(X)}_{\vec{a}',\delta_X'})\geq {\underline s}_{\vec{a}',\{\delta'_X\}_{X\in\ca{X}}}-{\underline\gamma}_{\vec{a},\{\delta'_X\}_{X\in\ca{X}},X},\nonumber
\end{align}
which leads to
\begin{align}
&\frac{1}{n}\log{D^{(X)}_{\vec{a}',\delta_X'}}-\frac{1}{n}\log{D^{(X)}_{\vec{a},\delta_X}}\nonumber\\
&\geq ({\underline s}_{\vec{a}',\{\delta'_X\}_{X\in\ca{X}}}-{\overline s}_{\vec{a},\{\delta_{X}\}_{X\in\ca{X}}})\nonumber\\
&-({\underline\gamma}_{\vec{a}',\{\delta'_X\}_{X\in\ca{X}},n}+{\overline\gamma}_{\vec{a},\{\delta_X\}_{X\in\ca{X}},X}).\label{eq:starplatinum}
\end{align}
Suppose ${\overline s}_{\vec{a},\{\delta_{X}\}_{X\in\ca{X}}}<{\underline s}_{\vec{a}',\{\delta'_X\}_{X\in\ca{X}}}$. Due to (\ref{eq:gammaconv}) and (\ref{eq:starplatinum}), we have
\begin{align}
\log{D^{(X)}_{\vec{a}',\delta_X'}}-\log{D^{(X)}_{\vec{a},\delta_X}}\geq0
\end{align}
for any sufficiently large $n$, which implies \eqref{keyofL12}. This completes the proof of the first part of Theorem \ref{Thm2}.

We prove the second statement by showing that an adiabatic transformation $(a,\{\delta_X\}_{X\in\ca{X}})\rightarrow_{ad}(a',\{\delta'_X\}_{X\in\ca{X}})$ is {\it not} possible if $\delta s:={\underline s}_{a',\{\delta'_X\}_{X\in\ca{X}}}-{\overline s}_{\vec{a},\{\delta_{X}\}_{X\in\ca{X}}}>0$. From \eqref{eq:boundpmmm}, we have
\begin{align}
&\frac{1}{n}\log{D^{(X)}_{\vec{a},\delta_X}}-\frac{1}{X}\log{D^{(X)}_{\vec{a}',\delta_X'}}
=\frac{1}{n}S(\hat{\pi}^{(X)}_{\vec{a},\delta_X})-\frac{1}{X}S(\hat{\pi}^{(X)}_{\vec{a}',\delta_X'})\nonumber\\
&\ge({\underline s}_{\vec{a},\{\delta_{X}\}_{X\in\ca{X}}}-{\overline s}_{\vec{a}',\{\delta'_X\}_{X\in\ca{X}}})-({\underline\gamma}_{\vec{a},\{\delta_X\}_{X\in\ca{X}},n}+{\overline\gamma}_{\vec{a}',\{\delta'_X\}_{X\in\ca{X}},X})\nonumber\\
&=\delta s-\tilde{\gamma}_X
\end{align}
for any $X$, where we defined $\tilde{\gamma}_X:={\underline\gamma}_{\vec{a},\{\delta_X\}_{X\in\ca{X}},X}+{\overline\gamma}_{\vec{a}',\{\delta'_X\}_{X\in\ca{X}},X}$.
Hence we obtain
\begin{align}
\frac{1}{D^{(X)}_{\vec{a}',\delta'_{X}}}>\frac{e^{-X\delta_{s}/2}}{D^{(X)}_{\vec{a}',\delta'_{X}}}\geq\frac{1}{D^{(X)}_{\vec{a},\delta_{X}}}\Label{eq:ddddc}
\end{align}
for any sufficiently large $n$. Note that $\lim_{n\rightarrow\infty}\tilde{\gamma}_X=0$ from \eqref{eq:gammaconv}. Therefore, as proved in Section \ref{s5-1}, we have \eqref{eq:mikitaka} for any sufficiently large $n$ and any unital CPTP map $\ca{E}_{X}$. This contradicts Condition \eqref{eq:defgenad}, which implies that an adiabatic transformation $(a,\delta_{X})\rightarrow_{ad}(a',\delta_{X}')$ is not possible. \hfill$\blacksquare$


\begin{thebibliography}{00}
\bibitem{Fermi}E. Fermi, {\it{Thermodynamics}} (Dover Books on Physics, 1956).
\bibitem{Bardeen}J. M. Bardeen, B. Carter, S. W. Hawking, Comm. Math. Phys. \textbf{31}, 161 (1973).
\bibitem{Shimizu}A. Shimizu, {\it{Basis of thermodynamics}}, (University of Tokyo Press (in Japanese), ISBN: 4130626094, (2007).)
\bibitem{Lieb}E. H. Lieb and J. Yngvason Phys. Rep. \textbf{310}, 1, (1996).
\bibitem{tasakistatistical}H. Tasaki, \textit{Statistical mechanics 1,2}, ISBN-13: 978-4563024376 and ISBN-13: 978-4563024383 (Baihukan, 2008 (in Japanese)).




\bibitem{Landau}Landau, L. D. and Lifshitz, E. M. {\it{Statistical Physics. Vol. 5 of the Course of Theoretical Physics}}, 3rd edition (1976).
\bibitem{Jarzynski}C. Jarzynski, Phys. Rev. Lett. \textbf{78}, 2690, (1997). 
\bibitem{tasaki}H. Tasaki, arXiv:cond-mat/0009244 (2000).
\bibitem{Kurchan}J. Kurchan, arXiv:cond-mat/0007360(2000). 
\bibitem{Car1}S. DeLiberato and M. Ueda, Phys. Rev. E \textbf{84}, 051122 (2011).
\bibitem{Xiao}G. Xiao and J. Gong, arXiv:1503.00784, (2015).
\bibitem{Sekimoto}K. Sekimoto, \textit{Stochastic Energetics (Lecture Notes in Physics)}, Springer, (2010).
\bibitem{tasaki15}H. Tasaki, arXiv:1511.01999 (2015).

\bibitem{Touchette}H. Touchette, Physics Reports \textbf{478}, 1 (2009).

\bibitem{Horodecki}M. Horodecki and J. Oppenheim, Nat. Commun. \textbf{4}, 2059 (2013).
\bibitem{oneshot1}O. C. O. Dahlsten, R. Renner, E. Rieper, and V. Vedral, New. J. Phys.\textbf{13}, 053015, (2011).  
\bibitem{oneshot3}J. Aberg, Nat. Commun. \textbf{4}, 1925 (2013).
\bibitem{Egloff}D. Egloff, O. C. O. Dahlsten, R. Renner and V. Vedral, arXiv:1207.0434, (2012).
\bibitem{Brandao}F. G. S. L. Brandao, M. Horodeck, N. H. Y. Ng, J. Oppenheim, and S. Wehner, PNAS, 112,3215(2015).
\bibitem{Car2}S. Popescu, arXiv:1009.2536.(2010).
\bibitem{Popescu2014}P. Skrzypczyk, A. J. Short and S. Popescu, Nature Communications \textbf{5}, 4185, (2014).
\bibitem{Tajima2014}H. Tajima and M. Hayashi  arXiv:1405.6457 (2014).
\bibitem{Weilenmann2015}M. Weilenmann, L. Kr\"{a}mer, P. Faist, and R. Renner, arXiv:1501.06920(2015).
\bibitem{review}G. Gour, M. P. Muller, V. Narasimhachar, R. W. Spekkens, N. Y. Halpern, arXiv:1309.6586 (2013).


\bibitem{openquantum}H. P. Breuer and F. Petruccione, {\it{The Theory of Open Quantum Systems}} (Oxford University Press, USA, 2007).
\bibitem{Nielsentext}M. A. Nielsen and I. L. Chuang, {\it{Quantum Computation and Quantum Information}} (Cambridge University Press, Cambridge, 2000).
\end{thebibliography}
\end{document}